\documentclass[10pt]{article}

\usepackage[english]{babel}

\usepackage{graphics,color}
\usepackage{epsfig}
\usepackage{amssymb}

\begin{document}
\date{}
\begin{center}
{\Large\bf Evolution of the quantized field coupled to a thermal bath: a phase space approach}
\end{center}
\begin{center}
{\normalsize E.P. Mattos and A. Vidiella-Barranco \footnote{vidiella@ifi.unicamp.br}}
\end{center}
\begin{center}
{\normalsize{ Gleb Wataghin Institute of Physics, University of Campinas - UNICAMP}}\\
{\normalsize{ 13083-859,   Campinas,  SP,  Brazil}}\\
\end{center}
\begin{abstract}
We present an alternative method for describing the evolution of a mode of the quantized electromagnetic 
field in contact with a finite temperature thermal bath. We employ the expansion of the field density 
operator in terms of coherent states and the related Glauber-Sudarshan $P$-function in phase space. 
The method allows us to obtain analytical expressions of the system's time-evolved $P$-function 
without needing to solve the corresponding master equation.  
\end{abstract}
\section{Introduction}

A density operator of a given quantum system may be represented in terms of distributions in phase space, 
known as quasiprobability distributions \cite{glauber69,wigner84}. Consider the case of a single mode of the 
electromagnetic field with creation and annihilation operators $\hat{a}^\dagger$ and $\hat{a}$, 
respectively. A family of quasiprobabilities ${\cal P}^{(l)}(\alpha)$ may be associated to a specific quantum state 
(having density operator $\hat{\rho}$) via the following complex Fourier transform of the 
characteristic function $\chi^{(l)}(\hat{\rho},\xi)$, or
\cite{glauber69} 
\begin{equation}
{\cal P}^{(l)}(\alpha) = \frac{1}{\pi^2}\int d^2\xi\, \exp\left( \alpha\xi^* - \alpha^*\xi\right) \chi^{(l)}(\hat{\rho},\xi).
\end{equation}
Here $\xi = \xi_r + i\xi_i$, $d^2\xi = d\xi_r d\xi_i$, and the integrations are performed over the whole
phase space (from $-\infty$ to $\infty$).
We may define the following three quasiprobability functions which are associated to characteristic functions having particular
orderings,

\noindent i) Normally ordered ($l = 1$):
$\chi^{(1)}(\hat{\rho},\xi) = Tr \left[\hat{\rho}\exp\left( \xi \hat{a}^\dagger\right) \exp\left( -\xi^* \hat{a}\right)\right]$ $\rightarrow$ ${\cal P}^{(1)}(\alpha) \equiv P(\alpha)$, the Glauber-Sudarshan 
$P$-function \cite{sudarshan63,glauber63}.

\noindent ii) Symmetrically ordered ($l = 0$):
$\chi^{(0)}(\hat{\rho},\xi) = Tr \left[\hat{\rho}\exp\left( \xi \hat{a}^\dagger-\xi^* \hat{a}\right)\right]$
$\rightarrow$ ${\cal P}^{(0)}(\alpha) \equiv W(\alpha)$, the Wigner function \cite{wigner32}.

\noindent iii) Anti-normally ordered ($l = -1$):
$\chi^{(-1)}(\hat{\rho},\xi) = Tr \left[\hat{\rho} \exp\left( -\xi^* \hat{a}\right)\exp\left( \xi \hat{a}^\dagger\right)\right]$ $\rightarrow$ ${\cal P}^{(-1)}(\alpha) \equiv Q(\alpha)$ the 
Husimi function or $Q$-function \cite{husimi40}.

Although in most cases the quasiprobabilities cannot be considered genuine joint probability distributions 
in phase space, they are functions containing all the information about the quantum state, i.e., they are equivalent 
to the density operator.
Quasiprobabilities have become very useful tools in quantum optics, e.g., they play a prominent role in quantum-state 
tomography \cite{risken89}, a scheme allowing the full reconstruction of the Wigner function via 
homodyne measurements \cite{raymer93}. They are also associated to nonclassicality criteria; for instance,
a quantum state is classified as ``classical" if its Glauber-Sudarshan $P$-function is a classical probability
density \cite{mandel86}. It is worth mentioning a measure of nonclassicality of quantum states based on the volume of 
the negative part of the Wigner function \cite{zyczkowski04}.
The $P$-function can be conveniently written as the weight function in the diagonal expansion of the density
operator of the field in the coherent state basis \footnote{A coherent state of the field $|\beta\rangle$ may be 
defined as $\hat{a} |\beta\rangle = \beta |\beta\rangle$, with $\beta \in \mathbb{C}$.}, i.e.,
\begin{equation}
\hat{\rho} = \int d^2\beta\, P(\beta) |\beta\rangle\langle\beta |.\label{pfunctionexpansion}
\end{equation}   
We remark that the $P$-function is singular for pure states. For example, from Eq. (\ref{pfunctionexpansion})
we note that for a coherent state $|\alpha\rangle$, $P(\beta) = \delta^{(2)}(\alpha - \beta)$. 
Interestingly, in spite of its singular character, it is possible to experimentally determine a nonclassical 
$P$-function, as shown in \cite{bellini08}.

Regarding the dynamics of quantum states, quasiprobabilities are particularly useful in transforming master 
(operator) equations into c-number equations. The Husimi ($Q$) function may be used for that purpose, for instance,
as for a given density operator we may write \cite{glauber69}
\begin{equation}
Q(\alpha) = \frac{1}{\pi} \langle \alpha|\hat{\rho}|\alpha\rangle,\label{qfunction}
\end{equation}
being $|\alpha\rangle$ coherent states. Thus, for a master equation of the form 
\begin{equation}
\frac{d\hat{\rho}}{dt} = \hat{L} \hat{\rho},\label{mastereqgeneral}
\end{equation}
a conversion to a c-number equation is possible if we act with a coherent state bra (and ket) 
in the following way
\begin{equation}
\langle \alpha|\frac{d\hat{\rho}}{dt}|\alpha\rangle = \pi \frac{dQ(\alpha)}{dt} 
= \langle \alpha|\hat{L} \hat{\rho}|\alpha\rangle.
\end{equation}
The resulting equation is often a Fokker-Planck type equation for the $Q$-function of the system.

In this contribution we propose a way to use the $P$-function to directly describe the evolution of 
the quantum state of a single mode field in contact with a finite temperature bath, instead of 
having to solve the corresponding master equation for a given initial quantum field state. In other
words, the process of solving an operator (or Fokker-Planck) equation is replaced by the solution
of integrals in phase space. As a result, we are able to obtain the time evolved $P$-function of the 
single mode field, as well as to calculate expectation values of the system operators in a straightforward 
way. This paper is organized as follows: in Section 2 we develop our method and provide some 
examples, and in Section 3 we present our concluding remarks. 

\section{The phase space method}

The evolution of the reduced density operator $\hat{\rho}$ of a single mode field in a dissipative cavity, i.e., in 
contact with a finite temperature bath in the Born-Markov approximation is given by the standard (interaction picture) 
master equation \cite{louisell73}
\begin{equation}
\frac{d\hat{\rho}}{dt} = \hat{L}(\Gamma,\overline{n}) \hat{\rho},\label{mastereqthermal}
\end{equation}
where the superoperator $\hat{L}(\Gamma,\overline{n})$ is such that
\begin{equation}
\hat{L}(\Gamma,\overline{n}) \hat{\rho}= 
\Gamma(1+\overline{n})\left(2\hat{a} \hat{\rho}\hat{a}^\dagger  - \hat{a}^\dagger \hat{a} \hat{\rho}
- \hat{\rho}\hat{a}^\dagger \hat{a} \right) + \Gamma\overline{n}\left(2\hat{a}^\dagger \hat{\rho}\hat{a}  - \hat{a} 
\hat{a}^\dagger \hat{\rho} - \hat{\rho}\hat{a} \hat{a}^\dagger \right).
\label{superopL}
\end{equation}
The parameter $\Gamma$ is the characteristic decay constant, and 
$\overline{n} = \left[\exp(\hbar\omega/k_B T)-1\right]^{-1}$ is the average number of excitations of the field mode 
having frequency $\omega$ at an effective temperature $T$.
Except in specific cases, e.g., for the field initially in a coherent state $|\alpha_0\rangle$ and the bath 
at $T = 0$K \cite{barnettbook}, it is not straightforward to find analytical solutions of the master equation above. 
As we have already mentioned, a possible approach is to convert the (operator) master equation into a c-number equation 
involving quasiprobabilities, for instance, the Husimi ($Q$-function) defined in Eq. (\ref{qfunction})\cite{hyuga96}.
Thus, the master equation Eq. \ref{mastereqthermal} can be transformed into the following equation
\begin{equation}
\frac{\partial Q(\alpha;t)}{\partial t} = \Big[\Gamma \Big(\frac{\partial}{\partial \alpha}\alpha +
\frac{\partial}{\partial \alpha^*}\alpha^*\Big) + 2\Gamma (1 + \overline{n}) 
\frac{\partial^2}{\partial\alpha\partial\alpha^*}\Big] Q(\alpha;t),
\end{equation}
i.e., the master (operator) equation has been replaced by a (c-number) Fokker-Plank equation.
   
Here we are going to adopt a different approach, based on the Glauber-Sudarshan $P$ representation. Firstly, we note that 
the formal solution of the master equation (\ref{mastereqthermal}) may be written as
\begin{equation}
\hat{\rho}(t) = e^{\hat{L}(\Gamma,\overline{n}) t}\hat{\rho}(0).
\end{equation}
Replacing the initial state $\hat{\rho}(0)$ by its $P$ representation, we obtain
\begin{equation}
\hat{\rho}(t) = e^{\hat{L}(\Gamma,\overline{n})t} \int d^2\beta\, P(\beta;0) |\beta\rangle\langle\beta | =
\int d^2\beta\, P(\beta;0) e^{\hat{L}(\Gamma,\overline{n})t} |\beta\rangle\langle\beta |.
\end{equation}
The term $e^{\hat{L}(\Gamma,\overline{n})t} |\beta\rangle\langle\beta |\equiv \hat{\rho}'(\beta,\overline{n};t)$ 
in the integrand corresponds exactly to the evolved density operator having the field mode initially in the 
coherent state $|\beta\rangle$. As discussed in \cite{hyuga96}, the solution of the corresponding master 
equation can be expressed in terms of displaced thermal states, or
\begin{equation}
\hat{\rho}'(\beta,\overline{n};t) = \frac{1}{1+\overline{n}_t}\sum_k \left(\frac{\overline{n}_t}{1+\overline{n}_t}\right)^k 
\hat{D}(\beta_t)|k\rangle\langle k|\hat{D}^\dag(\beta_t),
\end{equation}
where $\beta_t = \beta\ e^{-\Gamma t}$, $\overline{n}_t = \overline{n}(1-e^{-2\Gamma t})$, $\hat{D}(\beta_t) = 
\exp(\beta_t \hat{a}^\dagger - \beta^*_t \hat{a})$ is a time-dependent Glauber's displacement operator, 
and $| k\rangle$ are Fock states. 

Therefore, the field density operator at a time $t$ for an initial state having
a $P$-function $P(\beta;0)$ is given by 
\begin{equation}
\hat{\rho}(t) = \int d^2\beta\, P(\beta;0)\hat{\rho}'(\beta,\overline{n};t).\label{densityoperatort}
\end{equation} 

\subsection{Time evolution of physical quantities}

If we are interested in investigating the evolution of physical quantities, we would like to calculate the 
(time dependent) expectation values of the operators associated to the field, which can be done in a direct way. 
For an operator $\hat{O}$, we have that
\begin{equation}
\langle\hat{O}\rangle(t) = \int d^2\beta\, P(\beta;0)\  Tr[\hat{O}\ \hat{\rho}'(\beta, \overline{n};t)].
\end{equation}
We firstly calculate the trace $Tr[\hat{O}\ \hat{\rho}'(\beta, \overline{n};t)]$ and perform the integral 
weighted by $P(\beta)$, the $P$-function of the initial state. For the annihilation operator  $\hat{a}$, 
\begin{equation}
Tr[\hat{a}\ \hat{\rho}'(\beta, \overline{n};t)] = \beta_t,
\end{equation} 
so that
\begin{equation}
\langle\hat{a}\rangle(t) = \int d^2\beta\, P(\beta;0)\ \beta_t = e^{-\Gamma t}\ \int d^2\beta\, P(\beta;0)\ \beta = 
\langle\hat{a}\rangle(0)\ e^{-\Gamma t}.
\end{equation}
Now for the number operator $\hat{n} = \hat{a}^\dagger \hat{a}$,
\begin{equation}
\langle\hat{a}^\dag\hat{a}\rangle(t) = \int d^2\beta\, P(\beta;0)\ (|\beta_t|^2 + \overline{n}_t) = 
\int d^2\beta\, P(\beta;0)\ (|\beta|^2e^{-2\Gamma t} + \overline{n}_t) = 
\langle\hat{a}^\dag\hat{a}\rangle(0)e^{-2\Gamma t} + \overline{n}_t.
\end{equation}
We may thus verify how the thermal noise coming from the bath can affect the statistical properties of the 
cavity field, by calculating the evolution of quantities such as Mandel's ${\cal Q}$ parameter, for instance
\begin{equation}
{\cal Q}(t) = \frac{\langle \left(\Delta\hat{n}\right)^2\rangle - \langle\hat{n}\rangle}{\langle\hat{n}\rangle} 
= \frac{[\langle\hat{a}^{\dag2}\hat{a}^2\rangle(0) - \langle\hat{a}^\dag\hat{a}\rangle^2(0)]e^{-4\Gamma t} 
+ 2\overline{n}_t\langle\hat{a}^\dag\hat{a}\rangle(0)e^{-2\Gamma t} 
+ \overline{n}_t^2}{\langle\hat{a}^\dag\hat{a}\rangle(0)e^{-2\Gamma t} + \overline{n}_t}.
\end{equation}
Besides, the squeezing properties of the field can be evaluated by calculating the variances of the field 
quadrature operators $\hat{X} = (\hat{a} + \hat{a}^\dag)/2$
and $\hat{Y} = (\hat{a} - \hat{a}^\dag)/2i$:
\begin{equation}
\langle(\Delta \hat{X})^2\rangle(t) = \frac{2\overline{n}_t + 1}{4} + \left[\langle(\Delta \hat{X})^2\rangle(0) 
- \frac{1}{4}\right]\ e^{-2\Gamma t},
\end{equation}
with an analogous expression for $\langle(\Delta \hat{Y})^2\rangle(t)$.

\subsection{Time evolution of the quantum state: $P$-function}

This procedure also allows the calculation of the time-evolved Glauber-Sudarshan $P$-function of the cavity field.
Consider an initial field state $\hat{\rho}(0)$ having a $P$-function $P(\beta;0)$, or
\begin{equation}
\hat{\rho}(0) = \int d^2\beta\, P(\beta;0) |\beta\rangle\langle\beta |.
\end{equation}
We can express the displaced state $\hat{\rho}'(\alpha,\overline{n};t)$ in terms of its $P$-function 
$P'(\alpha,\gamma,\overline{n};t)$ as
\begin{equation}
\hat{\rho}'(\gamma,\overline{n};t) = \int d^2\alpha\, P'(\alpha,\gamma,\overline{n};t) |\alpha\rangle\langle\alpha |.
\end{equation}
Now, using Eq. (\ref{densityoperatort}) the field density operator at time $t$ can be written as
\begin{equation}
\hat{\rho}(t) = \int d^2\beta\, P(\beta;0)\int d^2\alpha\, P'(\alpha,\beta,\overline{n};t) |\alpha\rangle\langle\alpha | \equiv \int d^2\alpha\, P(\alpha;t) |\alpha\rangle\langle\alpha |.
\end{equation}
Thus, the system's time dependent $P$-function, $P(\alpha;t)$, is given by
\begin{equation}
P(\alpha;t)=\int d^2\beta\, P(\beta;0) P'(\alpha, \beta, \overline{n};t).
\end{equation}
Using the following result for the displaced thermal state, 
\begin{equation}
P'(\alpha, \beta, \overline{n};t) = \frac{1}{\pi \overline{n}_t} \ e^{-|\alpha-\beta_t|^2/\overline{n}_t},
\end{equation}
we finally obtain 
\begin{equation}
	P(\alpha; t)=\frac{1}{\pi \overline{n}_t}\int d^2\beta\,\ P(\beta; 0)\ e^{-|\alpha-\beta_t|^2/\overline{n}_t}.
	\label{finallyPfunction}
\end{equation}
We therefore conclude that given an initial state, $P(\beta; 0)$, it is possible to calculate the time-evolved 
$P$-function of the field mode by solving the phase-space integrals in Eq. (\ref{finallyPfunction}).

We would like to remark that in the particular case of having the bath at $T = 0$ K ($\overline{n} = 0$), 
a cavity field state 
initially in a coherent state $|\beta\rangle$ will remain
a coherent state, but with a decaying amplitude, i.e., $|\beta\rangle \rightarrow |\beta_t\rangle$ 
\cite{barnettbook}. 
As a consequence the time-evolved density operator corresponding to an initial state with $P$-function $P(\beta; 0)$
will be
\begin{equation}
\hat{\rho}(t) = e^{\hat{L}(\Gamma,0)t} \int d^2\beta\, P(\beta;0) |\beta\rangle\langle\beta | =
\int d^2\beta\, P(\beta;0) |\beta_t\rangle\langle\beta_t |.
\end{equation}
Performing a change of variables, we obtain
\begin{equation}
\hat{\rho}(t) =
\int d^2\beta\, P(\beta e^{\Gamma t};0) e^{2\Gamma t}|\beta\rangle\langle\beta |.
\end{equation}
This means that in the case of a zero temperature bath, the $P$-function of the time-evolved state will be 
given simply by
\begin{equation}
P(\alpha;t) = P(\alpha e^{\Gamma t};0) e^{2\Gamma t}.
\end{equation}

\subsection{Examples}
In order to illustrate our method for a finite temperature bath, in what follows we will calculate
$P(\alpha; t)$ for some initial cavity field states.

\subsubsection{Photon added thermal state}

As an example of initial non-Gaussian state we may consider the photon-added thermal state, defined as 
$\hat{\rho}^{(pats)}(0) = \left[Tr\left(\hat{a}^\dagger \hat{\rho}_{th}\hat{a}\right)\right]^{-1} \hat{a}^\dagger \hat{\rho}_{th}\hat{a} $, 
being $\hat{\rho}_{th}$ a thermal state with mean photon number $\overline{m}$.
Its associated $P$-function is
\begin{equation}
P^{(pats)}(\alpha;0)=\frac{\overline{m}+1}{\pi\overline{m}^3}\ 
\left(|\alpha|^2-\frac{\overline{m}}{\overline{m}+1}\right) \ e^{-|\alpha|^2/\overline{m}}.
\end{equation}
After performing the integrations in phase space, we obtain the $P$-function at a time $t$, 
\begin{equation}
P^{(pats)}(\alpha;t) = \left[\frac{(\overline{m}+1)e^{-2\Gamma t}}{\pi(\overline{m}e^{-2\Gamma t}+\overline{n}_t)^3}
|\alpha|^2+\frac{\overline{n}_t-e^{-2\Gamma t}}{\pi(\overline{m}e^{-2\Gamma t}+\overline{n}_t)^2}\right]\ 
\exp\left(-\frac{|\alpha|^2}{\overline{m}e^{-2\Gamma t}+\overline{n}_t}\right).
\end{equation}

\subsubsection{Photon added coherent state}

Another possible initial non-Gaussian state, but having a singular $P$-function, is the
photon-added coherent state $\hat{\rho}^{(pacs)}(0) = 
\left[Tr\left(\hat{a}^\dagger |\beta\rangle\langle\beta |\hat{a}\right)\right]^{-1} 
\hat{a}^\dagger |\beta\rangle\langle\beta | \hat{a} $, where $|\beta\rangle$ is a coherent state. 
In this case the initial $P$-function is 
\begin{equation}
P^{(pacs)}(\alpha;0)=\frac{\exp{\left(|\alpha|^2-|\beta|^2\right)}}{|\beta|^2+1}\frac{\partial^2}{\partial\alpha\partial\alpha^*}
\delta^{(2)}(\alpha-\beta).
\end{equation}
After the integrations in phase space we obtain the following $P$-function for $t > 0$
\begin{equation}
P^{(pacs)}(\alpha;t) = \frac{1}{\pi\overline{n}_t(|\beta|^2+1)}\left[\left|
\frac{e^{-\Gamma t}}{\overline{n}_t}\alpha+\left(1-\frac{e^{-2\Gamma t}}{\overline{n}_t}\right)\beta\right|^2
+\left(1-\frac{e^{-2\Gamma t}}{\overline{n}_t}\right)\right]
\exp\left(-|\alpha-\beta e^{-\Gamma t}|^2/\overline{n}_t\right).
\end{equation}

\subsubsection{Squeezed coherent state}

Now we consider an initial Gaussian state, but having a highly singular $P$-function, i.e., 
the squeezed coherent state defined as
$|\beta, s\rangle=\hat{D}(\beta)\hat{S}(s)|0\rangle$. Here $\hat{D}(\beta)$ is Glauber's displacement operator
and $\hat{S}(s) = \exp\left[\frac{s}{2}(\hat{a}^2 - \hat{a}^\dagger{}^2)\right]$ the squeezing operator.
We have taken $s \in \mathbb{R}$ for simplicity.
The corresponding $P$-function at $t = 0$ is \(\cite{schleichbook}\)
\begin{eqnarray}
P^{(scs)}(\alpha;0)&=&\exp\left(\frac{1-s}{8s}\frac{\partial^2}{\partial\alpha_r^2} + \frac{s-1}{8}
\frac{\partial^2}{\partial\alpha_i^2}\right)\ \delta^{(2)}(\alpha-\beta) \\ \nonumber  
&=& \left[\sum_n\frac{1}{n!}\left(\frac{1-s}{8s}\right)^n\frac{\partial^{2n}}{\partial 
\alpha_r^{2n}}\delta(\alpha_r-\beta_r)\right]\left[\sum_m\frac{1}{m!}\left(\frac{s-1}{8}\right)^m\frac{\partial^{2m}}{\partial 
\alpha_i^{2m}}\delta(\alpha_i-\beta_i)\right],
\end{eqnarray}
with $\alpha = \alpha_r + i \alpha_i$ and $\beta = \beta_r + i \beta_i$.
Performing again the integrations in phase space, we obtain the $P$-function for $t > 0$
\begin{eqnarray}
P^{(scs)}(\alpha;t)&=& 
=\frac{e^{-|\alpha-\beta e^{-\Gamma t}|^2/\overline{n}_t}}{\pi\overline{n}_t}
\left[\sum_n\frac{1}{n!}4^n \left(\frac{1-s}{8s}\right)^n\left(\frac{e^{-2\Gamma t}}{\overline{n}_t}\right)^n 
U\left(-n, \frac{1}{2}, \frac{(\alpha_r-\beta_re^{-\Gamma t})^2}{\overline{n}_t}\right)\right] \\ \nonumber
&\times&\left[\sum_m\frac{1}{m!}4^m\left(\frac{s-1}{8}\right)^m\left(\frac{e^{-2\Gamma t}}{\overline{n}_t}\right)^m
U\left(-m, \frac{1}{2}, \frac{(\alpha_i-\beta_ie^{-\Gamma t})^2}{\overline{n}_t}\right)\right].
\end{eqnarray}
Here $U(a,b,x)$ is the confluent hypergeometric function of the second kind \cite{arfkenbook}.

\section{Concluding remarks}

We have addressed the problem of the evolution of a cavity field coupled to a thermal bath using the coherent 
state expansion of the density operator, that is, the Glauber-Sudarshan $P$ representation.  Thus, rather 
than having to solve an operator equation (or a Fokker-Planck equation), it suffices to perform integrals 
in phase space using the $P$-function of the initial quantum state of the system. Employing the solution 
of the master equation in Eq. (\ref{mastereqthermal}) for an initial coherent state $|\beta\rangle$, 
we were able to analytically obtain the $P$-function of the time-evolved quantum state of the system for 
different initial conditions. Also, time-dependent physical quantities can be readily calculated, as 
expectation values of operators may be evaluated in a straightforward way.
Although we have presented a method for a specific dynamics (field mode 
in a thermal bath), it might be possible to extend the procedure to other types of system/bath interactions. 
Analytical approaches, which naturally allow a deeper understanding of the behavior of elementary quantum 
systems, are surely opportune at a time when we are witnessing intense activity in the field of 
quantum technology \cite{avb17}.

\section*{Acknowledgements}

This work was supported by FAPESP 
(Funda\c c\~ao de Amparo \`a Pesquisa do Estado de S\~ao Paulo),
grant N${\textsuperscript{\underline{o}}}$ 2019/00291-1, Brazil.



\end{document}